\documentclass[twocolumn]{revtex4}

\usepackage{graphicx}

\begin{document}

\title{Collective beliefs and individual stubbornness in the dynamics of public debates
}
\author{Serge Galam}
\email{serge.galam@polytechnique.edu}
\affiliation{Centre de Recherche en \'Epist\'emologie Appliqu\'ee,\\
\'Ecole Polytechnique and CNRS, \\CREA, Boulevard Victor, 32,
75015 Paris, France}

\begin{abstract}
The combined effects of collective beliefs and individual stubbornness in the dynamic of a public debate are investigated using the Galam sequential probabilistic model of opinion dynamics. The study is focused on pair interactions for which the bias produced by collective beliefs is the decisive factor to win the debate. The current value of that bias is a fixed external parameter. It is a constant of the problem not given to change. In contrast, stubbornness is an individual property. It results from external ingredients, which are susceptible to be modified during the debate. More precisely, giving some beliefs we determine the required stubbornness to oppose its associated bias in the debate outcome.
The results shed a new and counter intuitive light on paradoxical outcomes of sensitive issues, which are discussed in the public. The case of the global warming issue is discussed.
\end{abstract}
 

\maketitle

\newpage

\section{Introduction}

Sociophysics is nowadays a main stream of research in physics \cite{book, stauffer-book, galam-review, fortunato-review}. It covers numerous topics of social sciences including opinion dynamics, which has attracted a great deal of research  \cite{sznajd,  redner-2, frank-voter,  espagnol,  herrmann, schneider, hetero, sousa, lambiotte-ausloos,   pierluigi, mariage,  martins1, martins2}. Indeed, public opinion is a key feature to determine which decisions should be taken by policy-makers in modern democratic societies. It puts the understanding of its underlining mechanisms at the top priority among current major challenges. Any progress could have drastic effects on the way to tackle sensitive issues to which the global world is confronted.

In most bimodal opinion models the dynamics follows some deterministic flow, whose direction is determined by the existence of two attractors separated by a tipping threshold. The initial public supports are thus instrumental to determine the eventual outcome of the associated public debate. For each opinion, the major issue is to start the debate from an initial support beyond its tipping threshold. In some cases, only one single attractor drives the opinion dynamics \cite{contra}.

While all discrete models are characterized by a specific update rule, it was sown that all of them belong to one single probabilistic sequential scheme \cite{unify}. This approach was given some credit in 2005 with the successful prediction of a highly improbable outcome of a political vote \cite{hetero}. Moreover the prediction was made several months ahead of the actual vote against the  prediction of all polls and analyses \cite{lehir}. 

The key factor to this prediction was the use of collective beliefs to account for possible local doubt of agents discussing the issue. For a group of even size at a tie, Galam model attributes the choice to each opinion with a respective probability $k$ and $(1-k)$. The value of $k$ is a fixed constant not susceptible to be modified since it is the result of a long accumulation of the cultural traits of a social group.

It embodies the collective beliefs of agents by producing a bias in a local group when the agents are having a doubt about which opinion to adopt. Using a local majority rule for individual opinion updates, a doubt arises at a tie for an even size group \cite{hetero}. However for pair groups majority rule does not apply and $k$ is the unique parameter, which determines the final outcome of the dynamics. 

In parallel, the existence of stubbornness is an individual feature, which results from external ingredients. Given some knowledge and information an agent get persuaded to hold the true choice and thus becomes inflexible, i.e., it is no longer open to an eventual opinion shift. However, in contrast to collective beliefs, the ingredients on which the inflexibility is built can be weakened or dismissed during the debate. Accordingly, the current fraction of inflexibles is an external parameter, which can be modified, although it is not easily done.

In this paper we focus on pair interactions to determine the proportion of inflexibles, which is needed to oppose the bias produced by collective beliefs. More precisely, giving some beliefs we determine the required stubbornness to oppose its associated bias in the debate outcome. Simultaneously, it is seen how to amplify that collective bias by producing inflexibles along the same direction. The challenge is for each opinion to succeed in locating the unique attractor of the dynamics above fifty percent. Then the debate spontaneously guarantees its victory even if starting from a very low initial support in the public. Associated conditions are determined quantitatively. 

The results shed a new and counter intuitive light on paradoxical outcomes of sensitive issues, which are discussed in the public. It provides hints on how to design winning strategies in public issues. The case of the global warming issue is discussed  \cite{georgia, autre, ipcc, models, complex, belief, evo}.

\section{Modeling opinion interactions given a socio-pyschological frame}

We study a public issue for which two competing opinions A and B are available. Each agent does hold one of them. For a group of $N$ agents, selecting randomly an agent at time $t$, $p_t$ denotes the probability that this agent shares opinion A and $(1-p_t)$ is the probability that it holds opinion B.

\subsection{Internal versus external effects}

We consider that two different mechanisms are driving the opinion individual making off. One is external and includes the media, the available information, personal values and education. It acts directly on the agent. The second mechanism is internal to the group. It embodies the outcome of discussions about the actual issue among the group agents. In reality, both mechanisms are active during a public debate but with different time scales. The external effect is episodic and irregular while the internal one is sequential and regular. Here we focus on the internal mechanism, i. e., the internal driving dynamics of the public debate, assuming the external effect has been activated once prior to the time the public debate is launched. 

We make the hypothesis that at some time $t=0$ the proportions  $p_0$ and $(1-p_0)$ of agents sharing respectively opinions A and B are the net result of the external effect. They can be evaluated using polls. At that time the internal mechanism is turned on while the first one is cut of. We then focus on the time evolution of the A and B proportions at a series of discrete times $t=1, 2, 3... n$ which are produced by the interactions among agents. Indeed, while holding a given opinion, agents argue among them to validate their own choice. The process is monitored by local discussion which involve a few agents, most often in pairs.

\subsection{Floater versus inflexible agents}

Two types of agents are involved in the present model. Most are characterized as floaters, which are open-minded. Given significant counter arguments, a floater is inclined to shift its current opinion from A to B and vice versa. In contrast, there exist agents who are stubborn. Whatever argument is given to them, they do stick to their current opinion. To avoid a derogatory terminology we denote a stubborn agent as inflexible.

Inflexibility is grounded on external ingredients related to the issue. It could be claims that one of the two competing opinions A and B has been proven to be correct scientifically, let us say, opinion A.  While the claims are available to every agent, only a fraction of them takes them for granted. These very agents become inflexibles since for them the issue has been settled with opinion A being the correct one.  On this basis, no argument from other agents neither their own belief could trigger an opinion change. It is an individual property.

However, these external ingredients can be modified during the debate at the macroscopic level with the publication of new results or the discovery of new facts.  It might then turn some inflexibles to become floaters and vice versa. 

The respective densities of stubbornness  $a$ and $b$ for each opinion A and B obey the constrains $0\leq a \leq 1$, $0\leq b \leq 1$ and $0\leq a+b \leq 1$ and can also be evaluated using polls. Therefore at time $t$, we have $p_t$ supporters of opinion A of which $a$ are stubborn and $(p_t-a)$ are floaters. For opinion B it is $(1-p_t)$ supporters $b$ are stubborn and $(1-p_t-b)$ floaters. Accordingly the total proportion of floaters is $(1-a-b)$. We also have $a\leq p_t \leq 1-b$.

\subsection{The role of beliefs on a free choice making}

In addition to have two kind of agents, floaters and inflexibles,  we introduce the existence of collective beliefs within the population. Shared by all agents, these beliefs may be heterogeneous and normally do not intervene in the internal discussion among agents. However, in case a group of discussing agents reach a local state of doubt about the issue at stake with as much convincing arguments for opinion A as for opinion B, the collective beliefs are evoked to dispel that doubt. The opinion A or B in adequacy with the beliefs is selected by the agents in the local doubting group.

To account for this background psychological effect we denote $k$ the probability that a doubt yields to select opinion A against B and $(1-k)$ for B against A. The value of $k$ is a function of the group psyscho-sociological  frame and varies from one group to another with $0\leq k \leq 1$. It is the average result of the distributed beliefs within the whole group.

In case of a political change, most beliefs favor to preserve the status quo in case of a doubt about the benefit of a change. Accordingly couple of people sharing different opinions are expected to vote for the This effect has been corroborated from a 2004 data analysis from the American presidential election to study the marriage gap, i.e. the difference in voting for Bush and Kerry between married and unmarried people \cite{mariage}.

\subsection{Monitoring the local opinion dynamics}

To implement the global dynamics driven by local discussions among agents we assume that all agents are distributed randomly by groups of size $L$, which is restricted here to $L=2$, i.e., only pair interactions are considered. Within each pair when both agents share the same opinion nothing happens with $A A   \rightarrow  A A$, $A a   \rightarrow  A a$, $a a  \rightarrow  a a$ and $B B   \rightarrow  B B$, $B b   \rightarrow  B b$, $b b  \rightarrow  b b$ where A and B denote stubborn agents holding opinion respectively A and B with $a$ and $b$ being the corresponding floaters.
Otherwise the agents are supporting opposite views. In this case several configurations occur with the following outcomes and the permuted ones :

\begin{equation}
\left.
\begin{array}{ccc}
A b   &   \rightarrow  A a  \\ \\
a B   &   \rightarrow  a B  \\ \\
a b   &   \rightarrow  a a 
\end{array} 
\right\} with  \; probability  \; k
\label{pair-k} 
\end{equation}
when at a tie collective beliefs favor opinion A and 

\begin{equation}
\left.
\begin{array}{ccc}
A b   &   \rightarrow  A b  \\ \\
a B   &   \rightarrow  b B  \\ \\
a b   &   \rightarrow  b b
\end{array} 
\right\} with  \; probability  \; (1-k)
\label{pair-(1-k)} 
\end{equation}
when at a tie collective beliefs favor opinion B. Last case

\begin{equation}
A B   \rightarrow  A B
\label{pair-0} 
\end{equation}
occurs at a tie where nothing happens due to mutual stubbornness of both agents no matter the collective beliefs.

Given a density $p_t$  at time $t$, calculating the probability of occurrence of above each configuration, we can evaluate the new density $p_{t+1}$ of agents sharing opinion A at time $(t+1)$ after one update as,

\begin{equation}
p_{t+1}= (1-2k) p_{t}^2 + \{1-(1-2k)-a(1-k)-b k \}p_t+a(1-k)  \ ,
\label{pair1} 
\end{equation}
which can be rewritten in the more convenient form,
\begin{equation}
p_{t+1}= \alpha p_{t}^2 + \{1-\alpha -\beta  -\gamma \}p_t+ \beta  \ ,
\label{pair2} 
\end{equation}
where $\alpha \equiv (1-2k), \beta \equiv a(1-k), \gamma \equiv b k$ with the associated constraints  $-1 \leq \alpha \leq 1, 0\leq  \beta \leq  1, 0\leq  \gamma \leq  1$.

\section{The drastic effect of pair interactions in opinion dynamics}

Assuming the external parameters $a, b, k$ are given we now study the opinion dynamics driven by a series of repeated updates monitored by successive redistributions of all agents in random pairs. Starting from an initial proportion $p_t=p_0$ at the initial time $t=0$ we get the series $p_0 \rightarrow p_{1} \rightarrow p_{2}... \rightarrow p_{n}$ after $n$ opinion updates where Eq. (\ref{pair2})  is used at each time.

The strategic question is to determine if the public debate reaches a stable state after a finite number $n$ of updates, which in turn becomes the democratic expression of the public about the discussed issue. As such it can used by policy makers to justify the implementation of new regulations. 

To investigate the dynamics obtained from Eq. (\ref{pair2}) we solve the associated fixed point Equation $p_{t+1}=p_t$, which yields the two solutions,
\begin{equation}
p_{1,2}=\frac{1}{2 \alpha } \Big\{\alpha +  \beta  +  \gamma
 \pm \sqrt {-4  \alpha  \beta + (\alpha+  \beta +\gamma)^2}  \Big\}\ ,
\label{p12} 
\end{equation}
provided $\alpha \neq 0$.

\subsection{The case of balanced collective beliefs}

When collective beliefs are balanced or are not activated, we have $k=1/2$, which is equivalent to $\alpha = 0$. Either both agents keep unchanged their opinions or both adopt either A or B with the same probability, which on average results in a null effect. From the ties, an equal number of floaters adopt respectively opinions A and B, making Eq. (\ref{pair2}) to become,
\begin{equation}
p_{t+1}=  \{1 -\beta  -\gamma \}p_t+ \beta  \ ,
\label{pair3} 
\end{equation}
which yields the fixed point Equation $p(a+b)=a$. 

When $a+b \neq 0$ there exists one single attractor $p^*=a/(a+b)$, meaning that initial conditions for floaters are irrelevant to determine the final outcome of the debate. If $a>b$, opinion A wins the debate and reciprocally for $a<b$. 

\begin{figure}
\includegraphics[width=.45\textwidth]{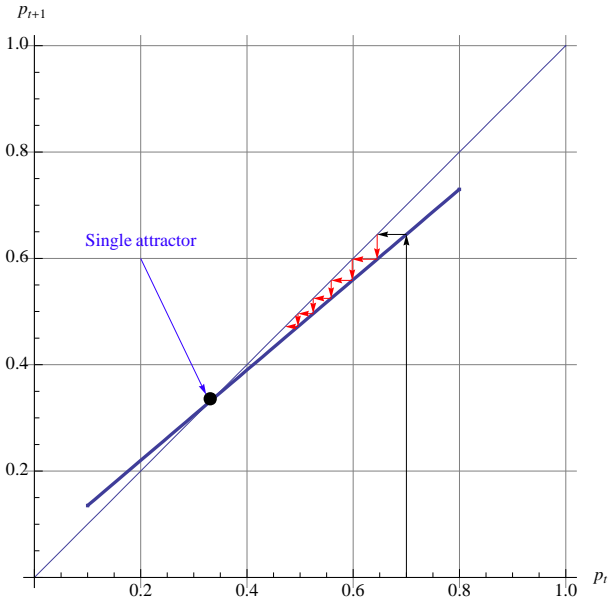}
\caption{The variation of $p_{t+1}$ is exhibited as a function of $p_{t}$ from $a=0.10$ to $1-b=0.80$ for $k=1/2, a=0.10, b=0.20$. There exists one single attractor $p^*=a/(a+b)=1/3<1/2$. An initial support of $70\%$ in favor of opinion A is found to end up with a victory of opinion B only after five updates. At the attractor only one third of the population is still holding opinion A. The case $k=1/2, a=b=0$ is also shown with the diagonal straight line  $p_{t+1}=p_t$ with $p$ varying from 0 to 1. There, every point is an attractor since the opinion proportions are invariant under the opinion dynamics process.}
\label{k=1/2}
\end{figure}

One case with $a=0.10$ and $b=0.20$ is exhibited in Figure (\ref{k=1/2}). The fixed point $p^*=a/(a+b)=1/3<1/2$ monitors the opinion dynamics. Any initial condition ends up to a A victory at a stable proportion of one third of agents in favor of A and two third in favor of B. The situation is illustrated with an initial support of $p_t=70\%$ in favor of opinion A, which is shown to shrink to one third with eventually a B victory. Only five updates are enough to reverse the majority with the series $p_0=0.70,  p_{1}= 0.645, p_{2}=0.598, p_{3}=0.558, p_{4}=0.525, p_{5}=0.496$. One more update accentuates the result with $p_{6}=0.472$.

Increasing the proportion of stubbornness on the A side by fifty percent at $a=0.15$ with still $b=0.20$ the associated series yields $p_{5}=0.532$ upholding now the A initial majority status. However three additional updates leads to $p_{8}=0.487$. On the contrary to start with a much lower support would not have changed the final result. What matters here is only the relative strength of stubbornness.

In the case of absence of stubbornness with $a=b=0$ Eq.  (\ref{pair3}) writes $p_{t+1}=p_{t}$ making any value of $p_t$ a fixed point. The dynamics does not modify the proportion of respective support for A and B with every agent being a floater. The associated diagonal straight line is also shown in  Figure (\ref{k=1/2}).

\subsection{Unbalanced collective beliefs}
\label{unba}

When a population shares heterogenous beliefs we have $k\neq 1/2 \Longleftrightarrow  \alpha \neq 0$ making Eq. (\ref{pair2}) to exhibit the two fixed points given by Eq.  (\ref{p12}). However to be valid, the solutions must obey $a\leq p_{1,2} \leq 1-b$ since they represent the proportions of each opinion with respective densities of stubbornness $a$ and $b$. We thus need to determine the ranges of $\alpha, \beta, \gamma$ for which this constraint is fulfilled for each one of the solutions. 

\subsubsection*{The $p_1$ solution}
Starting with $p_1$ it happens to be more practical to look first at $0\leq p_{1} \leq 1$ instead of $a\leq p_{1} \leq 1-b$.  Two cases are to discriminated:

(i) When $\alpha >0 \Longleftrightarrow 0\leq k < 1/2$ the constraint is violated with $p_{1} > 1$ in the range $k\neq 0$ and $b\neq 0$. However  $p_{1} = 1$ at $k=0$ or and at $b=0$. Three combinations are possible. The first one, $k=0$ and $b\neq 0$, is not possible since then $p_1=1>1-b$. The second one, $k\neq 0$ and $b=0$, is possible only when $k\leq k_a \equiv (1-a)/(2-a)$ for which $p_1=1\leq 1-b=1$. For $k> k_a$ we have $p_1=a(1-k)/(1-2k)>1$. The last case, $k=0$ and $b=0$, is also relevant since there $p_1=1\leq 1-b=1$. Therefore $p_1=1$ is a valid fixed point for $b=0$ in the range $0\leq k\leq k_a$  with the additional constraint $k_a<1/2$ due to $\alpha >0$, which implies $a>0$.

(ii) For  $\alpha <0 \Longleftrightarrow 1/2< k \leq 1$ we found $p_1 <0$ for $a \neq 0$ and $k\neq 1$ making it non physical in this range. On the other hand $p_1=0$ at $k=1$ or and at $a=0$. 
Three combinations are again possible. When $k=1$ and $a\neq 0$ it is dismissed as non physical since then $p_1=0<a$. The case $a=0$ and $k\neq 1$ is possible only for $k\geq k_b\equiv 1/(2-b)$ since otherwise $p_1=1+bk/(1-2k)<0$ when $k< k_b$. Last case $k=1$ and $a=0$ is also valid with $p_1=0\geq a=0$. The fixed point $p_1=0$ is thus valid at $a=0$ in the range $k_b \leq k\leq 1$ with the additional constraint $k_b>1/2$ due to $\alpha <0$, which requires $b>0$.

Joining above results we conclude that $p_1$ is a valid fixed point only when stubbornness  does  not exist on one side and within some limited range of collective beliefs, which yields three cases.

\begin{figure}
\includegraphics[width=.45\textwidth]{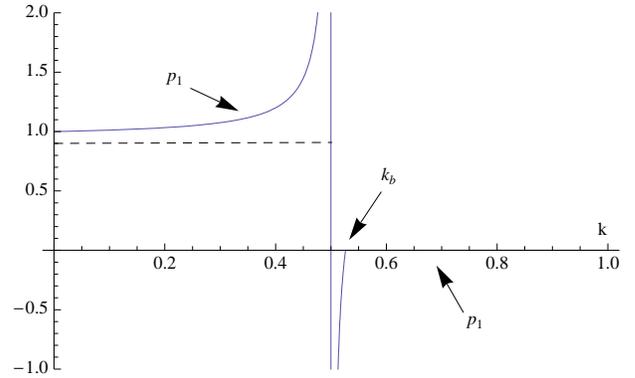}
\caption{The fixed point $p_{1}$ as a function of $k$ for $a=0$ and $b=0.10$. The part $p_1=1$ is not physical since $p_1$ must obey $p_1\leq 1-b=0.90$. The part $p_1=0$ is valid for $k_b \leq k\leq 1$ with $k_b\approx 0.526$ since we need to satisfy $p_1\geq a=0$.}
\label{p1b}
\end{figure}

\begin{figure}
\includegraphics[width=.45\textwidth]{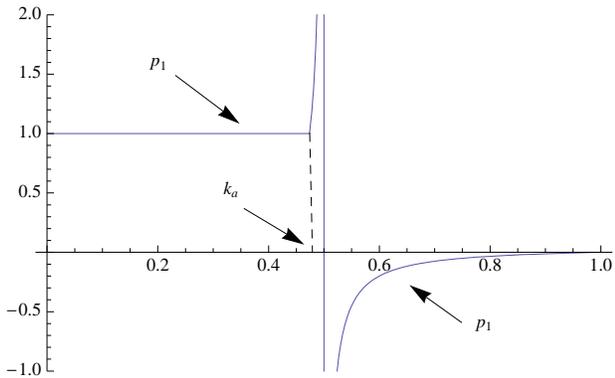}
\caption{The fixed point $p_{1}$ as a function of $k$ for $a=0.10$ and $b=0$. The part $p_1=1$ is valid only in the range $0\leq k\leq k_a\approx 0.474$ since $p_1$ must obey $p_1\leq 1-b=1$. The value $p_1=0$ at $k=1$ is not physical since we need to satisfy $p_1\geq a=0.1$.}
\label{p1a}
\end{figure}

\begin{description}
\item[(1)] For $a=0$, $p_1=0$ with collective beliefs restricted to the range $k_b \leq k\leq 1$ with $b>0$ where $k_b\approx 0.526$. Figure (\ref{p1b}) shows one case with $b=0.10$.

\item[(2)] For $b=0$ we have $p_1=1$ with collective beliefs being restricted to the range $0\leq k\leq k_a$ with $a>0$ where $k_a\approx 0.474$. Figure  (\ref{p1a}) shows one case with $a=0.10$.

\item[(3)] For any other values of $(a, b, k)$ the fixed point $p_1$ is non physical.

\end{description}

\begin{figure}
\includegraphics[width=.45\textwidth]{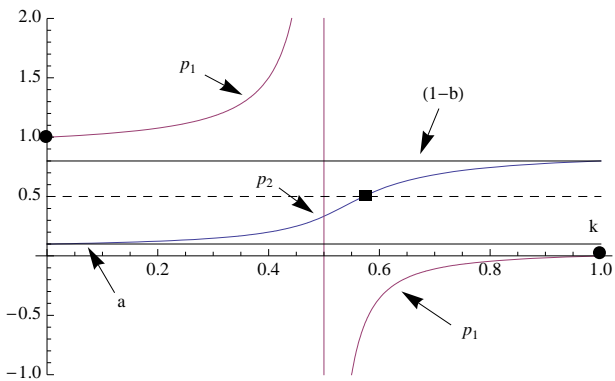}
\caption{The fixed points $p_{1,2}$ as a function of $k$ for $a=0.10$ and $b=0.20$. The $p_1$ solution is seen to be always unphysical since it  $p_1=1>1-b=0.90$ and $p_1=0<a=0.10$. The two extreme points at $k=0$ and $k=1$ are exhibited as full circles. In contrast $p_2$ satisfies $0.10\leq p_2\leq 0.90$ for the full range $0\leq k \leq 1$. The full square located on $p_2$ shows the value $k_c$ at which $p_2=1/2$, i.e., A wins the majority when $k>k_c$ and looses the debate for $k<k_c$.}
\label{roots1}
\end{figure}

While no stubbornness is required on one side, it is worth noticing that simultaneously at least an infinitesimal proportion of stubbornness is required on the other side. Otherwise  to have both $a=0$ and $b=0$ make $k_a=k_b=1/2$ which is exclude here. In the vicinity of no stubbornness with $a\approx 0$ and $b\approx 0$, we have $k_a\approx 1/2-a/4$ and $k_b\approx 1/2+b/4$. When $a\neq 0$ and $b\neq 0$ the fixed point $p_1$ is always non physical as can be seen in Figure (\ref{roots1}) where one case with  $a= 0.10$ and $b= 0.20 $.

\subsubsection*{The $p_2$ solution}

To be valid, $p_2$ given by Eq.  (\ref{p12}) must obey  $a\leq p_{2} \leq 1-b$. The first condition $p_{2} \leq 1-b$ is equivalent to $(1-k) b (1-a-b)\geq 0$, which is always satisfied since $(1-k)\geq 0$, $b \geq 0$ and $(1-a-b) \geq 0$. The equality $p_2=1-b$ is obtained for $k=1$ or and $b=0$ or and $a+b=1$. The other constraint $p_{2} \geq a$ yields $k a (1-a-b)\geq 0$. The conditions  $k=0$ or and $a=0$ or and $a+b=1$ yields $p_2=a$.  Last case $a=1-b$ makes true simultaneously both equalities $p_2=a$ and $p_2=1-b$. These extreme cases are considered in details in the next Section.

Contrary to $p_1$, which is valid only at peculiar sets of values for $(a, b, k)$, $p_2$ is always a valid fixed point. Indeed it is the unique attractor of the dynamics driven by Eq. (\ref{pair2}). This feature is rather important since it discards the initial conditions in the making of the opinion dynamics final outcome. The associated dynamics turns out to be a threshold-less dynamics ending at a stable public opinion characterized by $p_2$. One illustration of  the variation of $p_2$ as  a function of $k$ is shown in Figure (\ref{roots1}) for  $a= 0.10$ and $b= 0.20$ together with $p_1$.


\section{The extreme cases of beliefs and stubbornness}

We investigate the extreme cases with respect to both stubbornness and collective beliefs, i.e. we consider separately and combined the cases $a=0$ (no stubbornness on the A side), $b=0$ (no stubbornness on the A side), $a=b$ (equal stubbornness on both sides), $k=0$ (doubting yields to choose B) and $k=1$ (doubting yields to choose A).

\subsubsection*{Absence of stubbornness $a=b=0$}

\begin{figure}
\includegraphics[width=.45\textwidth]{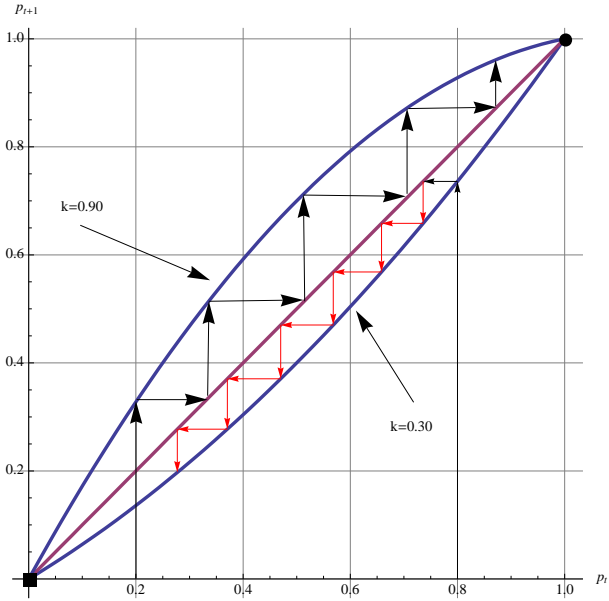}
\caption{The function $p_{t+1}$ as a function of points $p_{t}$ for $a=b=0$ with $k=0.30$ and $k=0.90$. In the first case, an initial very large support $p_0=0.80$ is seen to shrink down to $p_5=0.37$ and $p_7=0.20$. The attractor is located at zero noticed by a filled little square. In contrast a minority initial support $p_0=0.20$ is shown to be boosted  up to $p_5=0.96$ for $k=0.90$. The attractor is now located at one noticed by a filled little circle.}
\label{a=b=0}
\end{figure}

When $a=b=0$ the problem becomes rather simple.  Only the collective beliefs affect the opinion dynamics. Eq.  (\ref{pair1}) reduces to 
\begin{equation}
p_{t+1}= (1-2k) p_{t}^2 + 2k p_t \ ,
\label{pair4} 
\end{equation}
which has two fixed points $\bar{p}_1=0$ and $\bar{p}_2=1$ for $k\neq 1/2$. For  $k<1/2$ the attractor is $\bar{p}_1=0$ with $\bar{p}_2=1$ being an unstable fixed point. The dynamics is reversed at once when $k>1/2$ with $\bar{p}_1=0$ being the unstable fixed point towards the attractor located now at $\bar{p}_2=1$. An illustration is shown in Figure (\ref{a=b=0}) for the two cases $k=0.30$ and $k=0.90$.

In the first case, an initial very large support $p_0=0.80$ is found to shrink down to $p_5=0.37$ after five updates and to $p_7=0.20$ after two additional updates since the attractor is located at zero exhibited by a filled little square. In contrast a minority initial support $p_0=0.20$ is boosted  up to $p_5=0.96$ after five updates for $k=0.90$. The attractor is now located at one exhibited by a filled little circle.

Indeed, above description misses a mathematical singularity in the nature of the fixed points, which is revealed by increasing the $k$ one-dimensional parameter space to the three-dimensional parameter space $\{k, a, b\}$. The interplay between the two fixed points monitoring the dynamics happens to be subtle. 

\begin{figure}
\includegraphics[width=.45\textwidth]{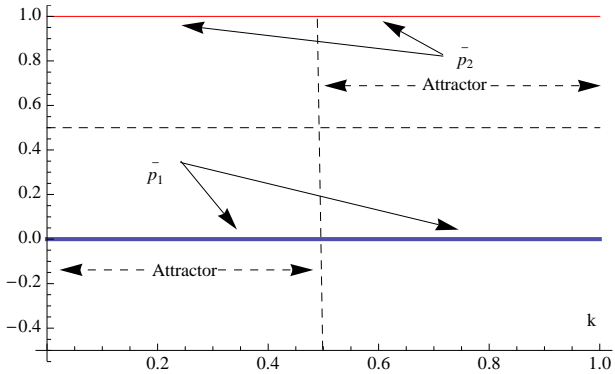}
\caption{The two solutions $p_1=0$ and $p_2=1$ obtained from Eq. (\ref{pair4}) where $p_1$ is the attractor for $k<1/2$ against $p_2$ when $k>1/2$}
\label{a=b=0-1}
\end{figure}

Using Eq. (\ref{pair4}) yields the two fixed points $\bar{p}_1=0$ and $\bar{p}_2=1$. Their stability analysis shows an exchange of stability with $p_1$ being the attractor for $k<1/2$ against $p_2$ when $k>1/2$ as exhibited in Figure (\ref{a=b=0-1}).

\begin{figure}
\includegraphics[width=.45\textwidth]{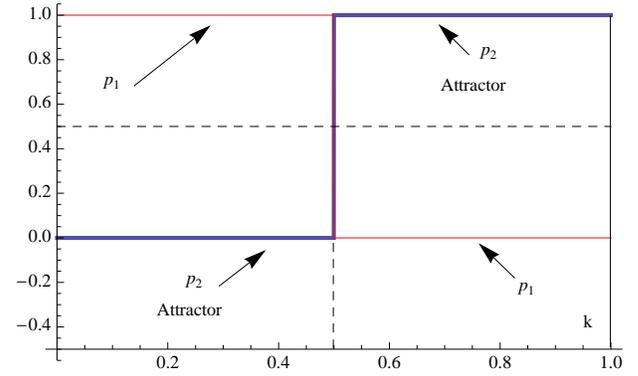}
\caption{The solutions from Eq. (\ref{p12}) where $a=b=0$ i.e., $\beta =\gamma=0$. A singularity is seen at $k=1/2$ where $p_1$ and $p_2$  exchange simultaneously their respective values with $p_1$ being always the attractor.}
\label{a=b=0-2}
\end{figure}

However plugging $a=b=0$ i.e., $\beta =\gamma=0$ into Eq. (\ref{p12}) reveals the existence of two subclasses of values for $p_1$ and $p_2$. Now $p_2$ is always the attractor but when $k<1/2$ we have $p_1=1$ and $p_2=0$ whereas $p_1=0$ and $p_2=1$ for $k>1/2$. The simultaneous exchange of values occurs at $k=1/2$. That singularity at $k=1/2$ corresponds in the $k$ one-dimensional plane to have all points invariant by the dynamics. The exchange of values is shown in Figure ( \ref{a=b=0-2}).

In the absence of stubbornness it is thus enough to defend an opinion which is just a little bit more consistent with the collective beliefs to be sure to win out the public debate provided an initial tiny cluster of floaters supports that opinion. In others words, for opinion A to have $p_0=0.01$ with $k=0.52$ guarantee to win the public debate. However this victory is somehow academic since many updates are required as shown by the series $p_{12}=0.02$ and $p_{40}=0.05$. In that case, an initial support of $p_0=0.35$  is required to reach the majority within 16 updates with $p_{16}=0.51$, which is still too much to be achieved with reasonable timescales of public debates. On the other hand, to take more advantage of the collective beliefs with $k=0.60$ ensure a strong victory within 4 updates with $p_4=0.54$. From $p_0=0.01$, 25 updates are required with $p_{25}=0.51$. But only seven creates a jump to $p_{7}=0.58$ with $k=0.95$. To take advantage of the collective beliefs is thus an incredible accelerator for minority spreading.

\subsubsection*{One-sided stubbornness $a=0$ with $b\neq 0$}

From above case, next step is to consider the case with one-sided inflexibles versus the effect of collective beliefs briefly discussed in Subsection \ref{unba}. Let us consider $a=0 \Longleftrightarrow \beta =0$ with $b\neq 0$.  Eq.  (\ref{pair1}) becomes 
\begin{equation}
p_{t+1}= (1-2k) p_{t}^2 + k(2-b)p_t \ ,
\label{pair5} 
\end{equation}
which has two fixed points $\bar{p}_1=0$ and $\bar{p}_2=1+\frac{b k}{1-2k}$ for $k\neq 1/2$. Last fixed point is physical only within the range $0\leq \bar{p}_2\leq (1-b)$, which requires both  $k> 1/2$ and $k\geq k_b$ where $k_b\equiv  \frac{1}{2-b}$, which reduces to $k\geq k_b$ with $b>0$ as found earlier. For  $0\leq k \leq k_b$, the unique valid fixed point $\bar{p}_1=0$ is the attractor. Whatever are the initial conditions, opinion A will eventually disappear. 

\begin{figure}
\includegraphics[width=.45\textwidth]{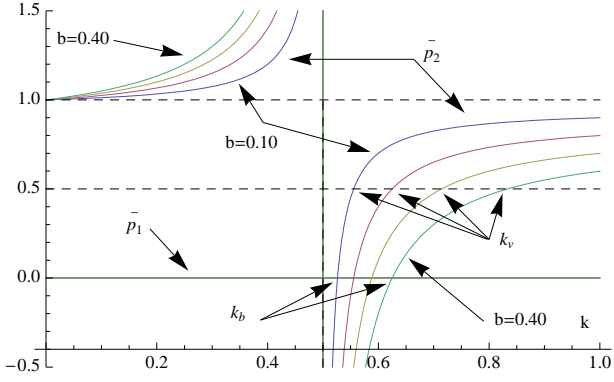}
\caption{The two solutions $\bar{p}_1=0$ and $\bar{p}_2=1+\frac{b k}{1-2k}$ from Eq. (\ref{pair5}) where $a=0$ i.e., $\beta =0$ and a fixed value of $b\neq 0$. Four cases are shown with $b=0.10, 0.20, 0.30, 0.40$.  For $k<k_b$ where $k_b\equiv  \frac{1}{2-b}$, $\bar{p}_1=0$ is the attractor and $p_2$ is non physical against $p_2$ as the attractor when $k>k_b$ with $p_1$ becoming unstable. Only for $k>k_v$ where $k_v\equiv  \frac{1}{2(1-b)}$ opinion A wins the debate.}
\label{b=0-1}
\end{figure}

In the range $k_b \leq k < k_v$, the two fixed points monitor the dynamics with now $\bar{p}_2$ being the attractor where $k_v\equiv  \frac{1}{2(1-b)}$. While opinion survives the debate it ends up as minority against opinion B, which wins the debate. At $k=k_v$ the two opinions stabilize at exactly fifty-fifty and only for $k_v < k \leq 1$ does opinion A wins the debate. Four cases are shown with $b=0.10, 0.20, 0.30, 0.40$ are shown in Figure  (\ref{b=0-1}).

\begin{figure}
\includegraphics[width=.45\textwidth]{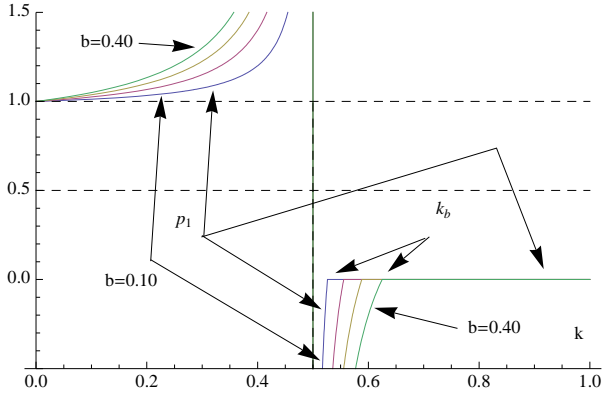}
\caption{The solution $p_1$ as a function of $k$ from Eq. (\ref{p12}) for $a=0$ i.e., $\beta =0$ and a fixed value of $b\neq 0$. Four cases are shown with $b=0.10, 0.20, 0.30, 0.40$. In the range $0<k<k_b$ where $k_b\equiv  \frac{1}{2-b}$, $p_1$ is not physical. At $k=0$ the value $p_1=1$ is a fixed point only for $b=0$. For $k_b\leq k\leq 1$,  $p_1=0$ is an unstable fixed point.}
\label{b=0-2}
\end{figure}

\begin{figure}
\includegraphics[width=.45\textwidth]{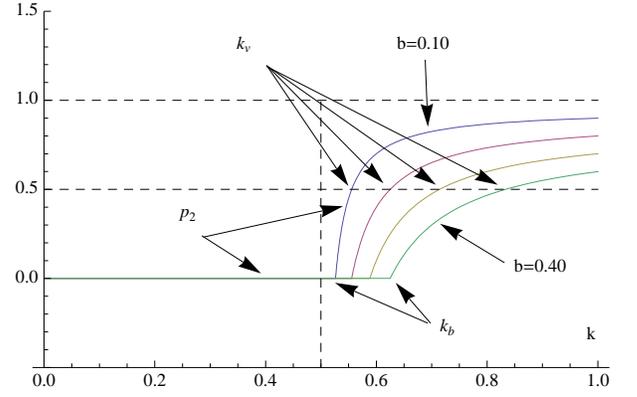}
\caption{The attractor $p_2$ as a function of $k$ from Eq. (\ref{p12}) where $a=0$ i.e., $\beta =0$ and a fixed value of $b\neq 0$. Four cases are shown with $b=0.10, 0.20, 0.30, 0.40$. In the range $0\leq k\leq k_b$ where $k_b\equiv  \frac{1}{2-b}$, $p_2=0$. For  $k_b<k<1$, $p_2 \neq 0$ with $p_2 >1/2$ when $k>k_v\equiv \frac{1}{2(1-b)}$. At $k=1$, $p_2=1-b$. }
\label{b=0-3}
\end{figure}

\begin{figure}
\includegraphics[width=.45\textwidth]{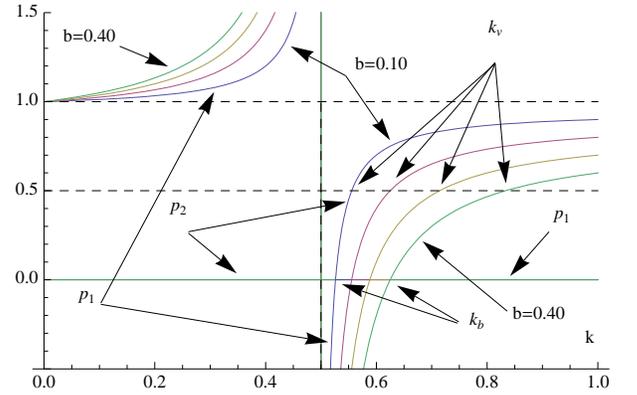}
\caption{The combination of both Figures  (\ref{b=0-2}) and (\ref{b=0-3}) showing the solutions $p_1$ and $p_2$ as a function of $k$ from Eq. (\ref{p12}) with $a=0$ i.e., $\beta =0$ and a fixed value of $b\neq 0$. Four cases are shown with $b=0.10, 0.20, 0.30, 0.40$. To be compared with Figures  (\ref{b=0-1}) where the solutions $\bar{p}_1$ and $\bar{p}_2$ from Eq. (\ref{pair5}). }
\label{b=0-4}
\end{figure}

However, starting from Eq.  (\ref{pair5}) hampers the visualization of the subtle interplay between the two fixed points obtained from Eq. (\ref{p12}) with $a=0$ as found in the precedent case $a=b=0$. At $k=0$ the fixed point $p_1=1$ is valid only for $b=0$ to satisfy $p_1\leq 1-b$ and $p_2=0$ is the attractor. For $0< k< k_b$ the solution $p_1$ is not physical with $p_2=0$ still being the attractor. At $k=k_b$ we have $p_1=p_2=0$. From $k> k_b$ the attractor $p_2$ turns non zero with $p_1=0$ becoming an unstable fixed point. At $k=1$ we have  $p_1=0$ and $p_2= 1-b$. Both results for $p_1$ and $p_2$ are exhibited in Figures  (\ref{b=0-2}) and  (\ref{b=0-3}) respectively for the four cases $b=0.10, 0.20, 0.30, 0.40$. The attractor $p_2$ turns fifty-fifty at $k=k_v$. For $k>k_v$ opinion A wins the debate. Figure  (\ref{b=0-4}) includes both fixed points and is to be compared with Figure  (\ref{b=0-1}).

\begin{figure}
\includegraphics[width=.45\textwidth]{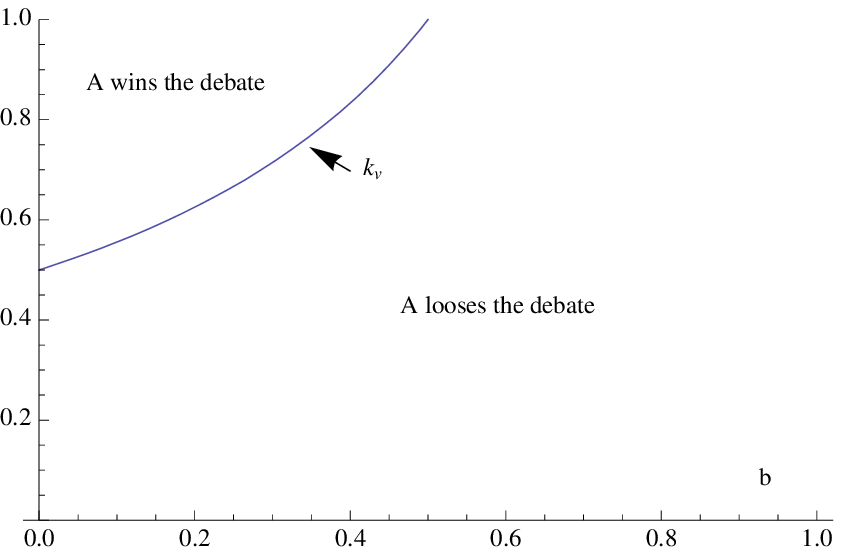}\\hfill
\includegraphics[width=.45\textwidth]{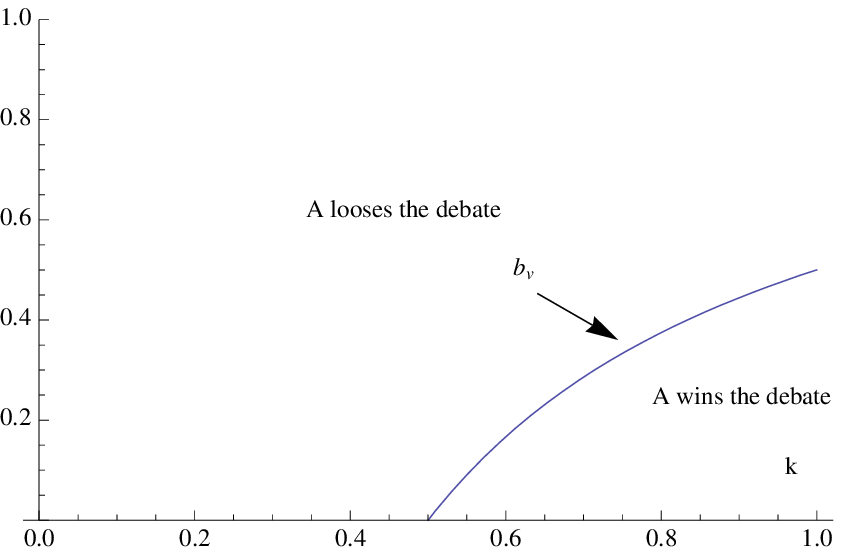}
\caption{Case $a=0$. Given $b$ the upper part shows the minimum value of collective beliefs required for A to win the debate with $k>k_v$ where $k_v\equiv 1/(2-2b)$. For $k<k_v$ A is defeated. Given $k$ the lower part shows the maximum value of B stubbornness which allows a A victory with $b<b_v$ where $b_v \equiv 1-\frac{1}{2k}$. A looses the debate for $b>b_v$.}
\label{kv-b}
\end{figure}

In summary, given a proportion $b$ of stubbornness on the B opinion side, a stubbornness free A opinion must benefit from a bias of the collective beliefs larger than $k_b$ not to be completely swept away. In order to win the debate, the A opinion must flow towards an attractor $p_2$ satisfying $p_2>1/2$, which is achieved for $k>k_v\equiv 1/(2-2b)$. The last constraint illustrates how to compensate stubbornness with collective beliefs and vice versa. Given a fixed value $k$, A looses the debate as soon as $b<b_v \equiv 1-\frac{1}{2k}$. The opinion landscape is shown in Figure (\ref{kv-b}). Only limited areas allow a A victory.

\subsubsection*{One-sided stubbornness $a\neq 0$ with $b= 0$}

\begin{figure}
\includegraphics[width=.45\textwidth]{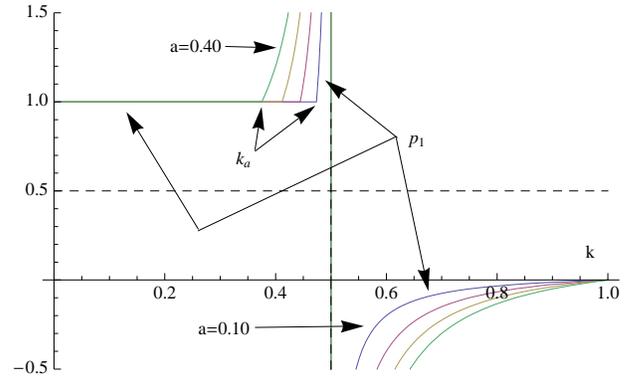}
\caption{The solution $p_1$ as a function of $k$ from Eq. (\ref{p12}) for $b=0$ i.e., $\gamma =0$ and a fixed value of $a\neq 0$. Four cases are shown with $a=0.10, 0.20, 0.30, 0.40$. In the range $0\leq k \leq k_a$ where $k_a\equiv 1- \frac{1}{2-a}$, $p_1=1$ is an unstable fixed point. For $k_a< k< 1$,  $p_1$ is not physical. The value $p_1=0$ at $k=1$ is valid only for $a=0$.}
\label{a=0-2}
\end{figure}

\begin{figure}
\includegraphics[width=.45\textwidth]{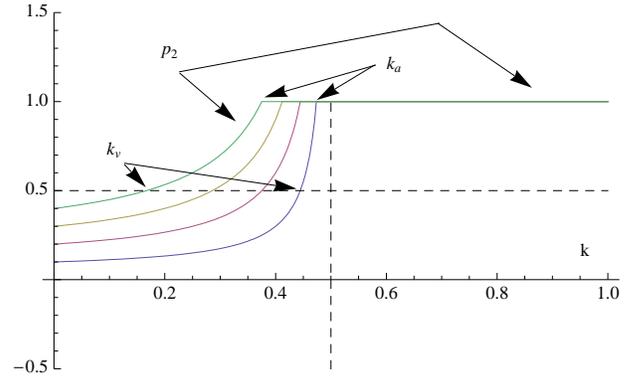}
\caption{The attractor $p_2$ as a function of $k$ from Eq. (\ref{p12}) where $b=0$ i.e., $\gamma =0$ and a fixed value of $a\neq 0$. Four cases are shown with $a=0.10, 0.20, 0.30, 0.40$. In the range $0\leq k < k_a$ where $k_a\equiv 1 -\frac{1}{2-a}$, $p_2\neq 1$. At $k=0$, $p_2=a$. For $0 \leq k< k_v$ where $k_v=1-\frac{1}{2(1-a)}$, $p_2<\frac{1}{2}$. At $k=k_v$ the attractor is exactly at fifty-fifty and only for $k>k_v$ does the A opinion wins the public debate with $p_2>\frac{1}{2}$. In the range $k_a \leq k \leq 1$, $p_2=1$ which means the B opinion will disappear in principle.}
\label{a=0-3}
\end{figure}

\begin{figure}
\includegraphics[width=.45\textwidth]{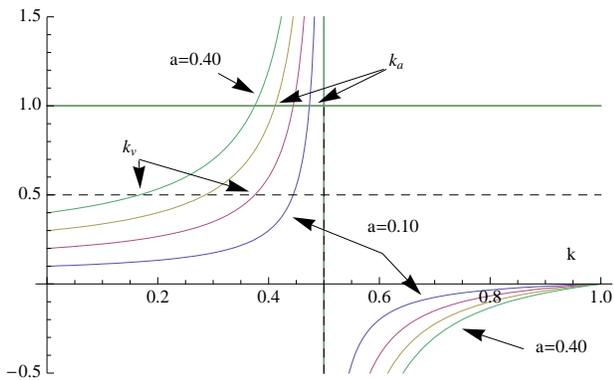}
\caption{The combination of both Figures  (\ref{a=0-2}) and (\ref{a=0-3}) showing the solutions $p_1$ and $p_2$ as a function of $k$ from Eq. (\ref{p12}) with $a\neq 0$ and $b=0$ i.e., $\gamma =0$. Four cases are shown with $a=0.10, 0.20, 0.30, 0.40$.}
\label{a=0-4}
\end{figure}

Since we are studying the topology of the dynamics landscape from the A opinion perspective it is worth to investigate the reverse situation where stubbornness is present only on the A side, i.e., $a\neq 0$ with $b= 0$. It is a symmetry change with respect to the precedent case $a=0$ i.e., $\beta =0$. However,  analyzing the equivalent of Figures  (\ref{b=0-2}, \ref{b=0-3}, \ref{b=0-4}) shown in Figures  (\ref{a=0-2}, \ref{a=0-3}, \ref{a=0-4}) sheds light on the appropriate strategy to be implemented by the A advocates in such a case. Only the results are given.

As seen from Figure (\ref{a=0-2}) the solution $p_1=1$ is a valid unstable fixed point in the range $0 \leq k\leq k_a$ where $k_a=1-\frac{1}{2-a}$. Otherwise it is non physical with $p_1>1$ for $k_a<k<\frac{1}{2}$ and $p_1<0$ for $\frac{1}{2}<k<1$, besides at $k=1$ when $a=0$ for which $p_1=0$. For $a=0.10, 0.20, 0.30, 0.40$ we have $k_a=0.47, 0.44, 0.41, 0.37$.

In contrast, the attractor $p_2$ is always valid. In the range $0 \leq k< k_v$ where $k_v=\frac{1-a}{2-a}$, $p_2<\frac{1}{2}$, which means the A opinion eventually looses the debate. At $k=0$, $p_2=a$. At $k=k_v$ the attractor is exactly at fifty-fifty and only for $k>k_v$ does the A opinion wins the public debate independently of the initial conditions as exhibited in Figure (\ref{a=0-3}). When $k_a \leq k\leq 1$ we have $p_2=1$, which means the B opinion will disappear in principle. 

The complete dynamics landscape is shown in Figure (\ref{a=0-4}). It combines both Figures (\ref{a=0-2}) and (\ref{a=0-3}).

\begin{figure}
\includegraphics[width=.45\textwidth]{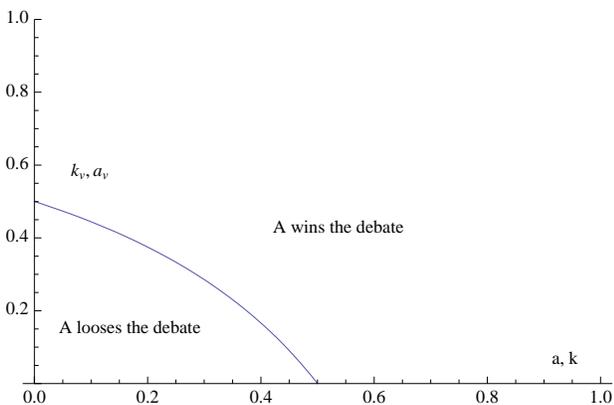}
\caption{Case $b=0$. Given a fixed value of $a$ ($k$) A wins the debate for $k>k_v$ ($a>a_v$) where $k_v\equiv 1/(2-2b)$ and $a_v\equiv 1/(2-2k)$.}
\label{kv-a}
\end{figure}

Given a proportion $a$ of stubbornness on its side, the A opinion can afford to be in part counter beliefs provided $k>k_v$. The larger $a$, the lower $k_v$ as shown in Figure (\ref{kv-a}). From $k_v=1-\frac{1}{2(1-a)}$ we get $a_v=1-\frac{1}{2(1-k)}$, which yields the minimum value of stubbornness required for opinion A to overcome a collective beliefs characterized with a fixed value $k<\frac{1}{2}$, i.e, mainly opposite to it. The dependences of $k_v$ on $a$ and of $a_v$ on $k$ are identical as illustrated in Figure (\ref{kv-a}). Given $a$ ($k$), we need $k>k_v$ ($a>a_v$) to have $p_2>1/2$. These constraints illustrate how to compensate stubbornness with collective beliefs and vice versa.

\subsubsection*{Extreme cases $k=0$ and $k=1$}

Plugging $k=0$ into Eq. (\ref{pair1}) reduces to,
\begin{equation}
p_{t+1}=  p_t^2-a p_t +a \ ,
\label{k=0} 
\end{equation}
which yields two fixed points located respectively at 1 and $a$ with the second one being the attractor. Every agent ends up sharing opinion B besides the proportion $a$ of inflexibles sharing opinion A.

At the other extreme with $k=1$ Eq. (\ref{pair1}) becomes,
\begin{equation}
p_{t+1}=  p_t^2(1-b) p_t  \ ,
\label{k=1} 
\end{equation}
which yields two fixed points located respectively at 0 and $(1-b)$ with the second one being the attractor. Every agent ends up sharing opinion A besides the proportion $b$ of inflexibles sharing opinion B.

\section{The general landscape of the winning opinion strategy}

\begin{figure}
\includegraphics[width=.45\textwidth]{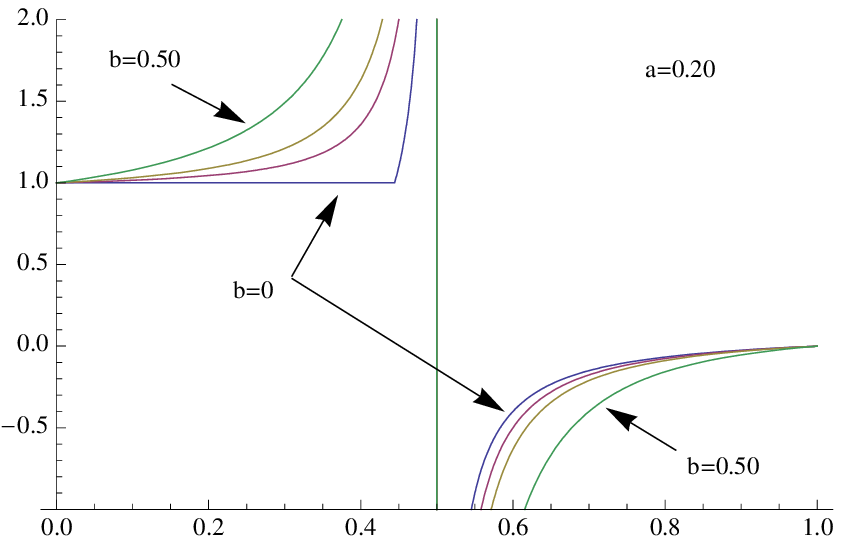}\\hfill
\includegraphics[width=.45\textwidth]{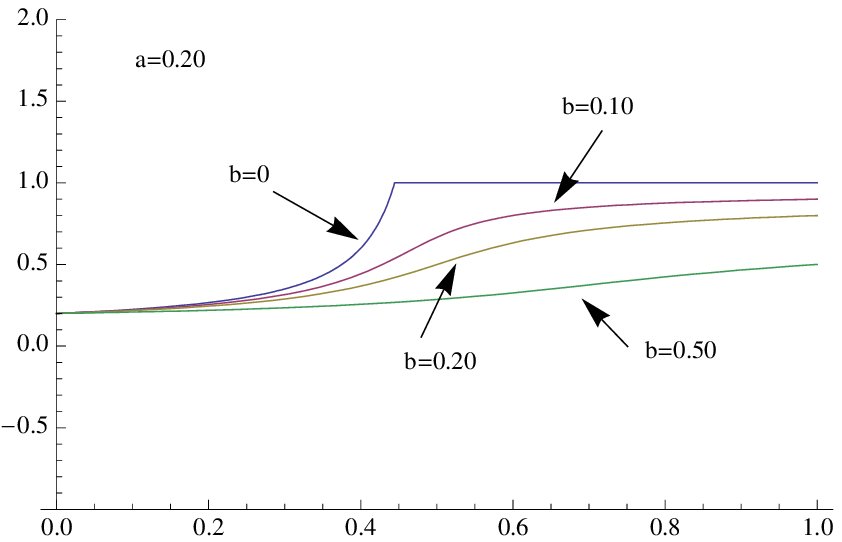}\\hfill
\includegraphics[width=.45\textwidth]{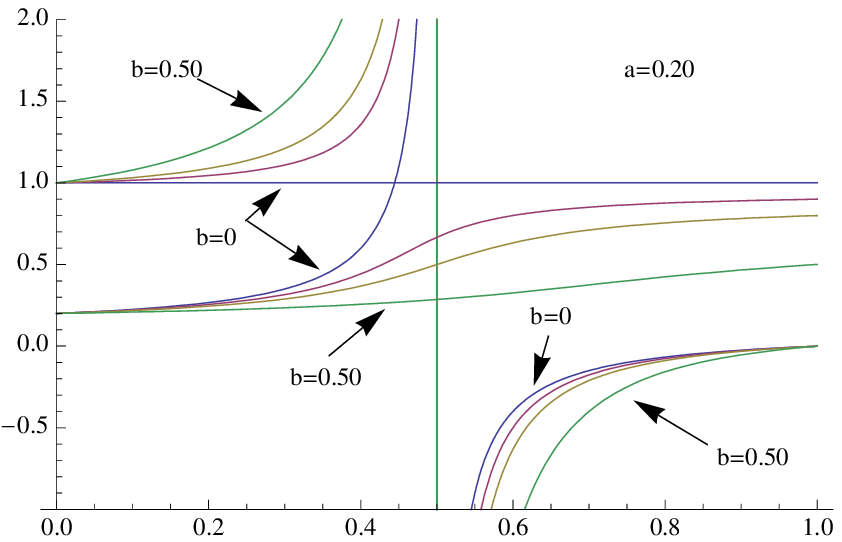}
\caption{Both functions $p_1$ and $p_2$ as a function of $k$ respectively for the 4 cases $b=0, 0.10, 0.20, 0.50$ with $a=0.20$. Only $p_1$ in the upper part, only  $p_2$ in the middle part and $p_1$ and $p_2$ in the lower part.}
\label{gen-b}
\end{figure}

\begin{figure}
\includegraphics[width=.45\textwidth]{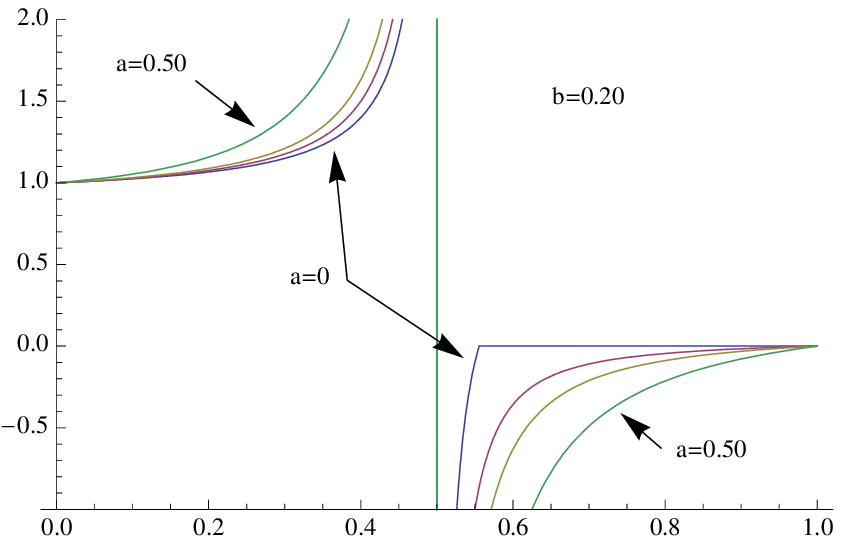}\\hfill
\includegraphics[width=.45\textwidth]{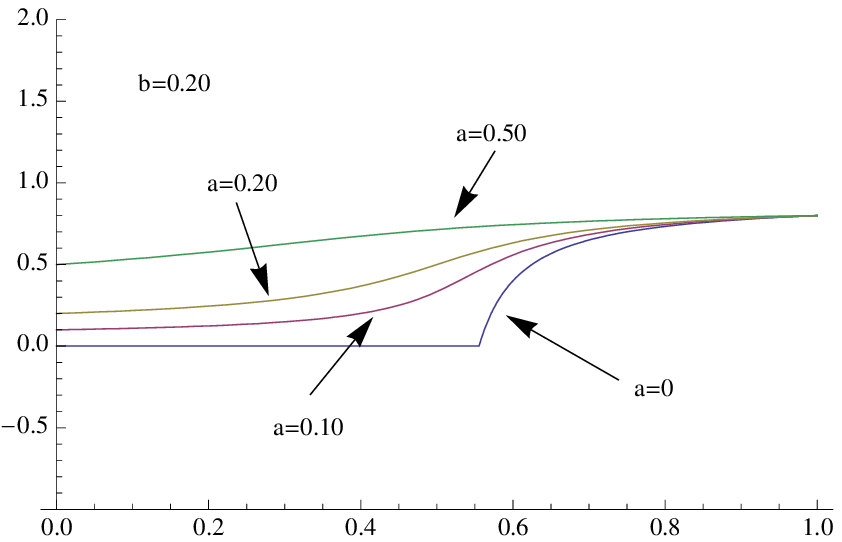}\\hfill
\includegraphics[width=.45\textwidth]{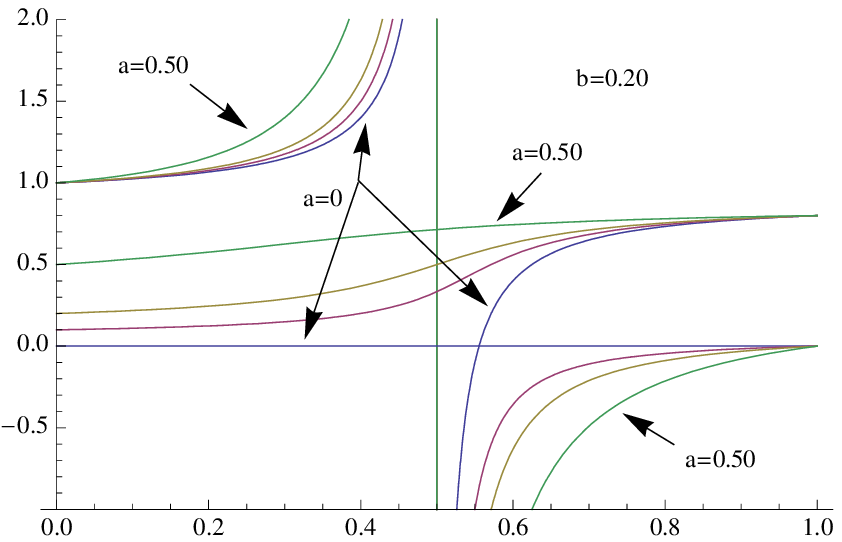}
\caption{Both functions $p_1$ and $p_2$ as a function of $k$ respectively for the 4 cases $a=0, 0.10, 0.20, 0.50$ with $b=0.20$. Only $p_1$ in the upper part, only  $p_2$ in the middle part and $p_1$ and $p_2$ in the lower part.}
\label{gen-a}
\end{figure}

We are now in a position to build the three dimensional $(k, a, b)$ landscape of the opinion dynamic flow monitored by the dependence of the two fixed points $p_1$ and $p_2$ on these three independent parameters. Knowing that $p_2$ is the attractor, the domain of validity of $p_1$ being indeed very limited, we can exhibit some two-dimensional plans. Figure (\ref{gen-b}) shows respectively the variation of $p_1$, $p_2$ and ($p_1, p_2$) as a function of $k$ for the 4 cases $b=0, 0.10, 0.20, 0.50$ with $a=0.20$. The symmetric situation with $a=0, 0.10, 0.20, 0.50$ with $b=0.20$ is exhibited in Figure (\ref{gen-a}).

\begin{figure}
\includegraphics[width=.45\textwidth]{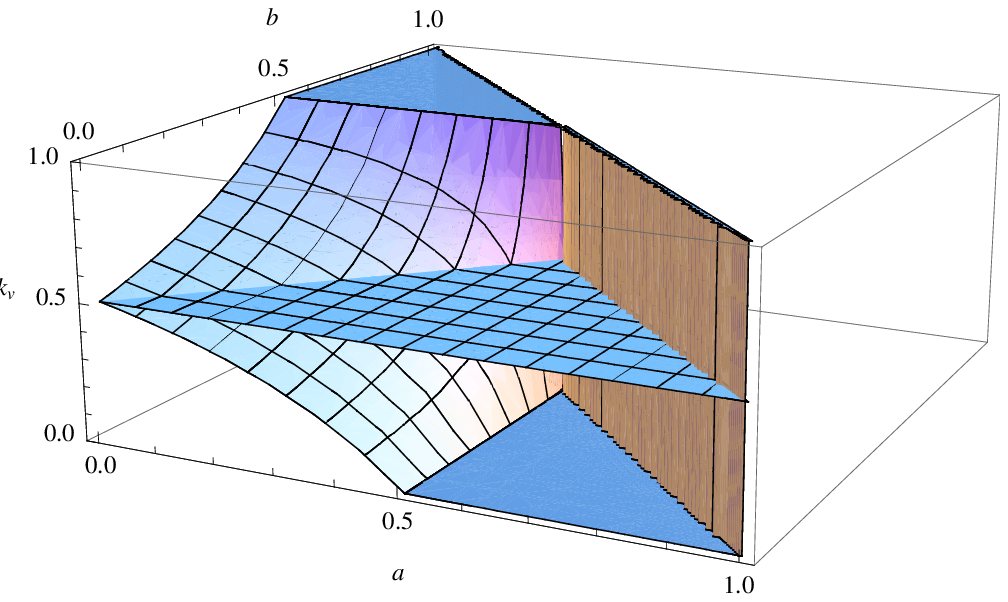}
\caption{The critical surface $k_{v}$ as a function of $a$ and $b$. When $k>k_{v}$ opinion A wins the debate. Otherwise it looses. The plane $k=1/2$ is also shown. Only the physical region  $a+b\leq 1$ is exhibited.}
\label{kab}
\end{figure}

In particular we can determine the critical surface where both opinions reach an attractor located at exactly fifty-fifty solving $p_t=1/2$ using Eq. (\ref{pair1}). It yields
\begin{equation}
k_{v}=  \frac{1-2a}{2(1-a-b)}  \ .
\label{kv} 
\end{equation}
This surface delimits both space for which opinion A respectively looses and wins the public debate independently of the initial proportions of floaters supporting A or B and is exhibited in Figure (\ref{kab}). 

\begin{figure}
\includegraphics[width=.45\textwidth]{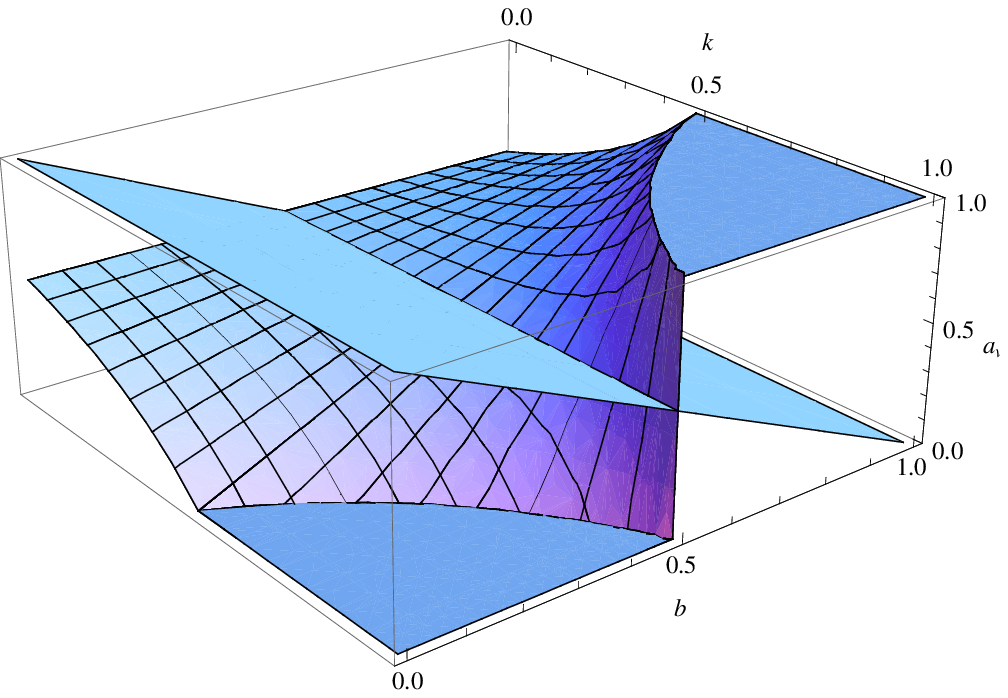}
\caption{The critical surface $a_{v}$ as a function of $k$ and $b$. When $a>a_{v}$ opinion A wins the debate. Otherwise it looses. The plane $a=1-b$ is also shown. Only the region where $a\leq 1-b $ is valid.}
\label{akb}
\end{figure}

However the time scale to modify collective beliefs is much longer than the one to build inflexibles, the instrumental formula is therefore to have $a$ as a function of both $k$ and $b$. It is obtained reversing Eq. (\ref{kv}) with
\begin{equation}
a_{v}=  \frac{1-2k+2b k}{2(1-k)}  \ .
\label{av} 
\end{equation}
It  is exhibited in Figure (\ref{akb}).

Both Figures (\ref{kab}) and (\ref{akb}) determine the geometry of the winning outcomes for a public debate between A and B, which are independent of the initial floater proportions. The practical highlight to eventually win out a public debate is to emphasize the fact that it is a single attractor dynamics. Accordingly, what matters in the conduct of a public debate from the A perspective is to achieve $p_2>1/2$. Then, for any initial support $p$, opinion A is certain to win no matters how low support A has at the beginning of the debate. Accordingly $p_2> 1/2$ when $k>k_v$ for fixed $a$ and $b$ and for $a>a_v$ for fixed $k$ and $b$.

The symmetric formula to determine the B best strategy is to have $b$ as a function of both $k$ and $a$ with
\begin{equation}
b_{v}=  \frac{-1+2k+2a(1-k)}{2k}  \ .
\label{bv} 
\end{equation}


\section {Application to the global warming issue}

Above mechanisms may enlighten some aspects of the current on going public debate about global warming, which is a public issue with an important controversial impact \cite{georgia, autre, ipcc, models, complex, belief, evo}. The debate articulates around two competing opinions. The first opinion (A) states that the recently observed global warming is man-made while the second opinion (B) dismisses that claim emphasizing that climatology is not yet a well established hard science. Opinion A holders assess that we are heading towards an unprecedented catastrophe in case we do not curb drastically green gas emissions. In contrast, opinion B supporters note that current climate is not worrying with respect to historical past situations.

In case of a doubt about the cause of the global warming, the general belief shared by a large majority of the public is that man activities are triggering  very negative impacts on our environment as well against the bio diversity. According most agents chose opinion A in case of non conclusive arguments about the actual cause of observed global warming.  The result of that situation is a value of the bias $k>1/2$ which favors opinion A. It means that in absence of inflexibles, the public debate drives inexorably the whole population towards an extremism state with everyone sharing the man made explanation.

In addition a substantial fraction of agents are convinced by the IPCC claims that  the man made cause is scientifically proved. Accordingly we have $a\neq 0$. In contrast B opinion holders are less inclined to be inflexibles since they are supporting a claim. They are only not convinced that A is true. It is also more difficult to convince someone to adopt B since it does not provide an an alternative view to A. We can thus infer that $b\neq 0$ but with the condition $b << a$. 

The two conditions  $k>1/2$ and $b << a$ makes impossible a victory of the B opinion.  To win the debate, i.e., to reach more than fifty percent support within the public, the B holders must satisfy $b>b_v$ since at $b=b_v$ the attractor locates precisely at fifty percent.  Otherwise the A opinion is victorious.

\begin{figure}
\includegraphics[width=.45\textwidth]{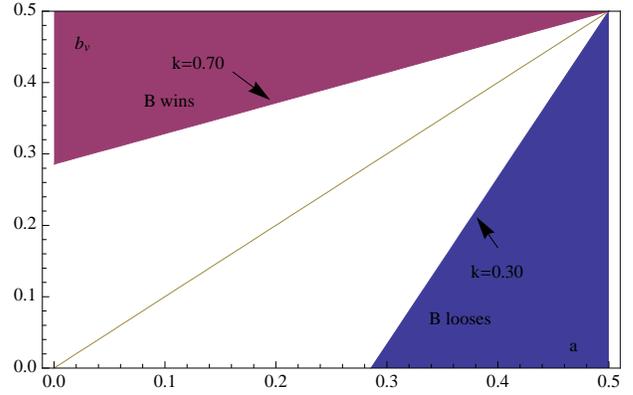}
\caption{The critical line $b_{v}$ as a function of $a$ for the two cases $k=0.70$ (favors A) and $k=0.30$ (favors B). When $k=0.70$ to overcome this bias in favor of A, B needs to build a rather large fraction of inflexibles within the upper grey triangle. Outside this area B is certain to loose the debate. The case  $k=0.30$, which favors B, yields a large area  for a B victory. Only within the lower grey triangle does B looses the debate.}
\label{global}
\end{figure}

Figure  \ref{global} shows the function $b_v$ as a function of the proportion $a$ of A inflexibles for  the two cases $k=0.70$ (favors A) and $k=0.30$ (favors B). When $k=0.70$ collective beliefs induce a bias in favor of opinion A. Then, to overcome this bias opinion B needs to build a rather large fraction of inflexibles as seen from the Figure. In addition it requires to satisfy the condition $b >> a$, which is totally unrealistic. 

The eventual reduction in A inflexibles is still without effect on the final outcome since at $a=0$ we have $b_v=0.29$. To win the debate, the B opinion must transform a large fraction of floaters into stubborn agents. Such a goal could be achieved by solving the question scientifically, which is far out of reach at the moment. However this result has to be tempered by the fact that in reality agents do not discuss only by pairs. Including larger sizes groups create the possibility to have a two attractor dynamics with a tipping threshold, which in turn makes possible to opinion B to overcome the collective beliefs biased against it.

In these conditions, only via a broad campaign among the public with large discussing groups, could opinion B take advantage of the recent repeated drawbacks suffered by the man made global warming holders. From the current reduction in public support towards opinion A, opinion B holders should keep on their activism to hold debates on the issue. Otherwise, we could infer that the recent A setbacks suffered by the man made global warming holders, which have resulted in a weakening of the public support, will be forgotten, and the driving force of the collective beliefs will lead again the public along a strong support of opinion A \cite{georgia, autre, ipcc, models, complex, belief, evo}.

\section{Conclusion}

Our results hint on how to design possible strategies to win a public debate. In particular it has shown how the existence of collective beliefs is a major factor in the direction taken by a public debate. To counter balance this strong bias, the making of inflexibles is efficient but requires a substantial fraction of them as seen from Figure  (\ref{global}).

Since the collective beliefs are not given to modifications within short timescales, the best approach for one opinion to win is to focus on getting as many as possible inflexibles along its side. However this goal could demand to overstate the validity of some arguments to sustain and legitimate that opinion.  In contrast, such a behavior could rise ethical questions. At the time it is of importance to prevent the formation of inflexibles on the other side by questioning the validity of the assessments used to promote it.

To conclude our analysis shed some new light on the possible laws, which govern the evolution of a public debate. It emphasize the role of global frame like the collective beliefs as well as individual features of the agents like stubbornness. 

Nevertheless, it should not be forgotten that we have dealing with a model, which aims to grasp some part of the reality, but  which is not the reality. We did not present a formal proof of some opinion dynamics but we have developed a novel angle to look at public debates, which could produce drastic changes in the design of lobby strategies.


\end{document}